\documentclass[showpacs,amsmath,amssymb,twocolumn]{revtex4}
\usepackage{pstricks,amsmath,bbm,mathrsfs,amssymb,psfrag,pifont,times,mathptmx}
\usepackage[utf8]{inputenc}
\usepackage{subfigure}
\usepackage[english]{babel}
\usepackage{color}
\usepackage{array}
\usepackage[draft]{graphicx}

\newcolumntype{P}[1]{>{\centering\arraybackslash}p{#1}}

\begin{document}
\title{Measuring nonclassicality with Silicon photomultipliers}
\author{Giovanni Chesi}
\affiliation{Department of Science and High Technology, University of Insubria, Via Valleggio 11, I-22100 Como (Italy)}
\author{Luca Malinverno}
\affiliation{Department of Science and High Technology, University of Insubria, Via Valleggio 11, I-22100 Como (Italy)}
\author{Alessia Allevi}
\affiliation{Department of Science and High Technology, University of Insubria, Via Valleggio 11, I-22100 Como (Italy)}
\author{Romualdo Santoro}
\affiliation{Department of Science and High Technology, University of Insubria, Via Valleggio 11, I-22100 Como (Italy)}
\author{Massimo Caccia}
\affiliation{Department of Science and High Technology, University of Insubria, Via Valleggio 11, I-22100 Como (Italy)}
\author{Maria Bondani}\email{maria.bondani@uninsubria.it}
\affiliation{Institute for Photonics and Nanotechnologies, CNR, Via Valleggio 11, I-22100 Como (Italy)}
\date{\today}




\begin{abstract}
Detector stochastic deviations from an ideal response can hamper the measurement of quantum properties of light especially in the mesoscopic regime where photon-number resolution is required. We demonstrate that, by a proper analysis of the output signal, nonclassicality of twin-beam states can be detected and exploited with commercial and cost effective silicon-based photon-number-resolving detectors.
\end{abstract}


\maketitle

\begin{figure}[htbp]
\begin{center}
\includegraphics[width=0.7\columnwidth]{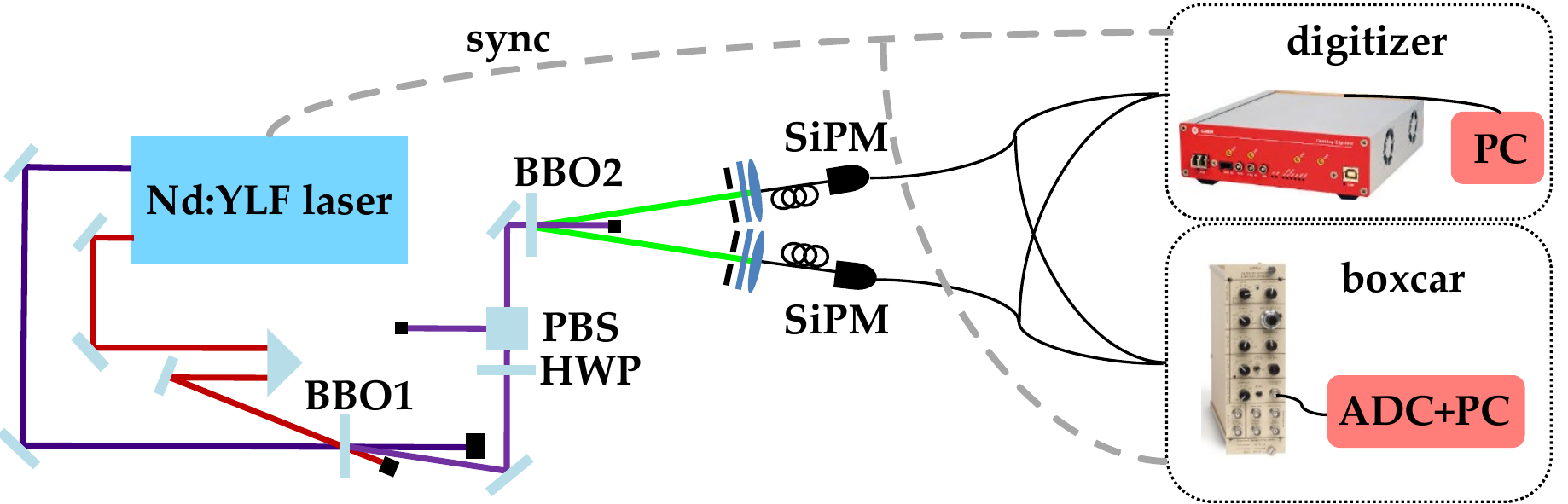}
\end{center}
\caption{Experimental setup for the measurement of multi-mode TWB states.}
\label{setup_TWB}
\end{figure}

\begin{figure}[h!]
\begin{center}
\includegraphics[width=0.7\columnwidth]{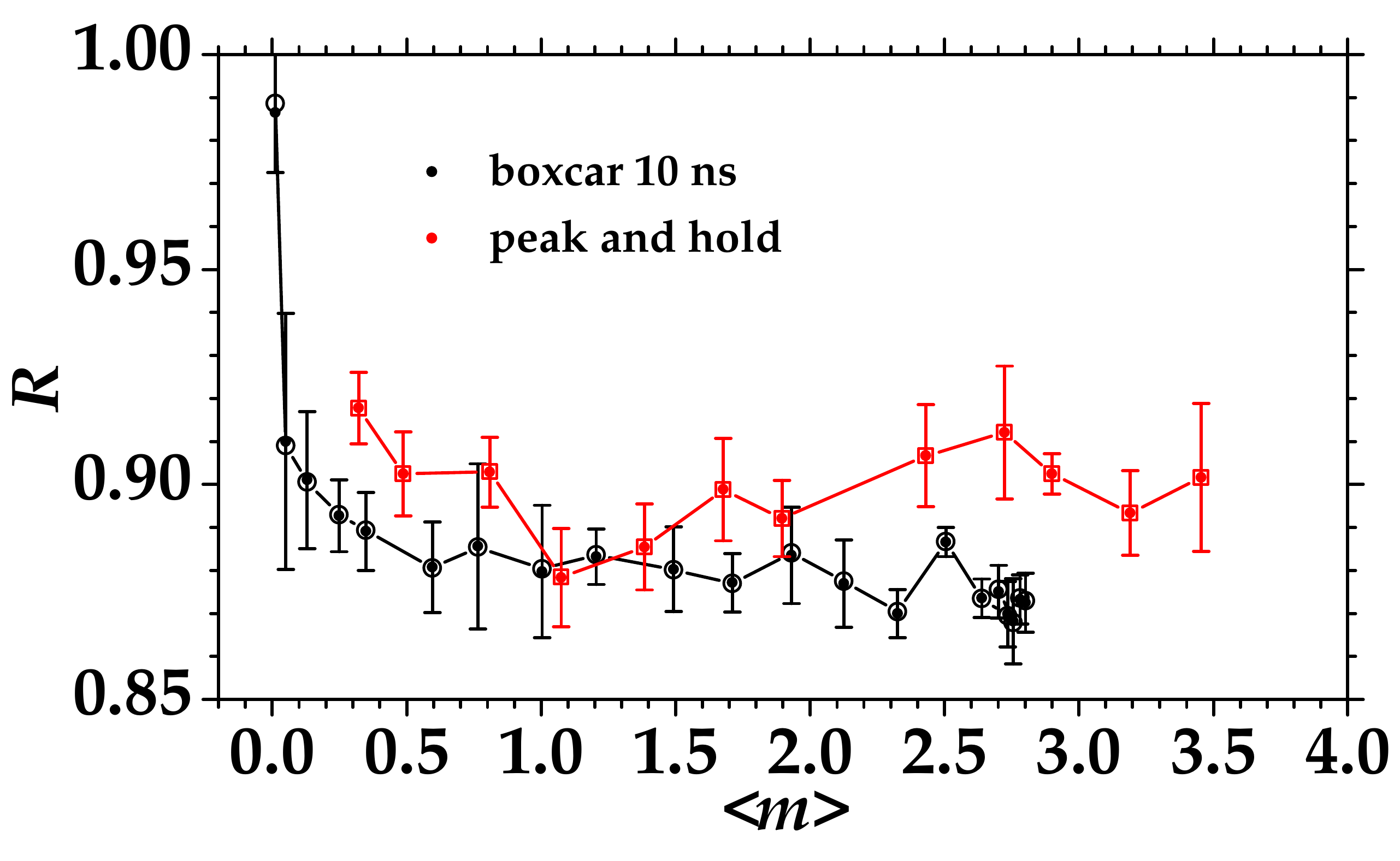}
\end{center}
\caption{Noise reduction factor as a function of the number of photons measured in a single arm: peak-and-hold measurement (red) compared to the results for minimum boxcar gate (10~ns, black). Full symbols are the experimental data and line+open symbol are the theoretical expectations.}
\label{peak}
\end{figure}

Quantum features of optical states are often fragile and their measurement requires advanced detection schemes to achieve sufficient quantum efficiency, low noise, no spurious effects, and photon-number resolution. This holds both in the case of single-photon states \cite{migdall} and in the case of mesoscopic states \cite{PRA07,arimondo}.\\
This is why quantum optics has triggered the research on new detectors, especially in the past twenty years \cite{lita,goltsman,rehacek,fitch}. The best performing detectors so far are quite complicated \cite{ave10} and require cryogenic operation \cite{harder16,bohmann}, even if photon-number-resolving (PNR) detectors, such as hybrid photodetectors \cite{JMO} have been successfully exploited for quantum measurements \cite{PRA12,PRA13}. Among PNR detectors, Silicon photomultipliers (SiPM, also known as multi-pixel photon counters), introduced in late '90s \cite{akindinov,bondarenko,saveliev,piemonte,renker}, feature photon-number resolution, wide dynamic range, reasonable quantum efficiency (up to 60$\%$ in the visible spectral range) and room temperature operability, that make them potentially attractive for quantum optics applications. Nevertheless, the presence of spurious effects have so far \cite{afek,kala12} prevented their exploitation in quantum optics. SiPMs consist of a matrix of avalanche photodiodes connected in parallel to a single output. Each diode is reverse-biased at a voltage value exceeding the breakdown threshold and it works in  Geiger M$\ddot{\rm u}$ller regime, yielding a standard output signal at any detection event. The number of active cells results in an output signal proportional to the overall number of detected photons.
In addition to the signal generated by the impinging light, the output also includes dark counts, cross talks and afterpulses. The dark-count rate can be decreased by lowering the temperature, while cross talk and afterpulse probabilities mainly depend on the detector architecture and technology. In the present generation of SiPMs, the cross talk probability has been lowered down to few percents and the afterpulses are almost negligible \cite{Hama1,Hama2}.\\
\begin{figure}[htbp]
\begin{center}
\includegraphics[width=0.7\columnwidth]{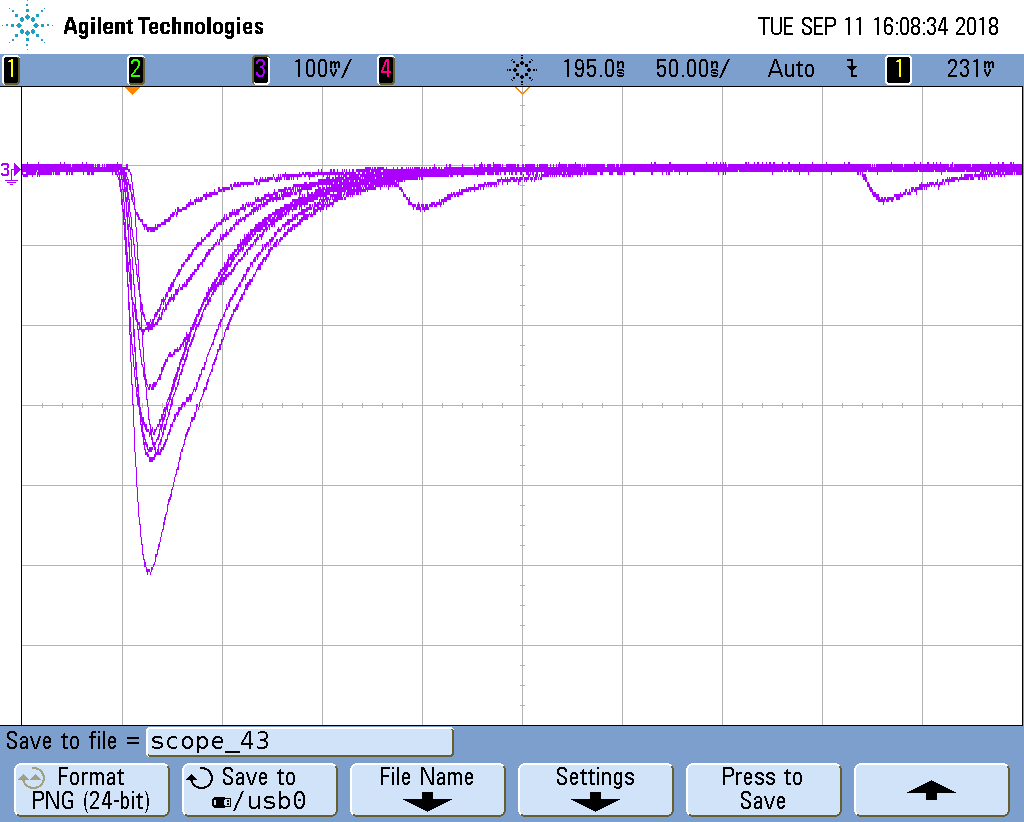}
\end{center}
\caption{Typical SiPM output in which delayed spurious events are visible.}
\label{output}
\end{figure}
\noindent
In spite of the high quality of the detectors, the residual imperfections can still severely affect the measurement of quantum features of the optical states. To minimize their influence we can take advantage of the different temporal occurrence of the spurious effects with respect to the light signal. Dark counts are randomly distributed in time, while cross-talk and afterpulse events are triggered by the detection events and thus can occur at a delayed time with respect to the primary detection. As an example, in Fig.~\ref{output} we show a typical SiPM response containing some delayed events. By properly choosing the temporal acquisition gate centered around the peak synchronous with the light pulses, we can get rid of the contribution of non-synchronous events. A detailed discussion is presented in a different paper \cite{manuscriptARXIV}.\\
Here we show the results obtained by measuring the same nonclassical states with the same detectors but changing the acquisition chain.\\
We consider the experimental setup in Fig.~\ref{setup_TWB}, in which mesoscopic twin beam (TWB) states are generated by parametric downconversion (PDC) in type-I quasi-collinear interaction geometry. The pumping field is the fourth harmonics of a Nd:YLF laser (4.5~ps pulse duration and 500~Hz repetition rate), while the nonlinear material is a $\beta$-barium-borate (BBO) crystal (BBO2, cut angle = 46.7 deg, 6-mm long). The generated TWB state is intrinsically multimode \cite{EPL} and the photon-number statistics of each of the two parties is well described by a multi-mode thermal distribution. Two portions of TWB around frequency degeneracy (523 nm), spatially and spectrally filtered by means of two variable irises and two bandpass filters, are delivered to the sensors via two multi-mode optical fibers (600-$\mu$m core diameter). In the experimental conditions under investigation, the effective number of independent thermal modes is about 100, so that the photon statistics closely resembles a Poissonian distribution, with a Fano factor $F=\sigma^2(n)/\langle n\rangle\simeq 1$.
The state characterization is performed as a function of the pump intensity, which is modified through a half-wave plate (HWP) followed by a polarizing cube beam splitter (PBS). For each energy value, $10^5$ single-shot acquisitions are performed by using two SiPMs (MPPC S13360-1350CS, Hamamatsu Photonics \cite{Hama1,Hama2}).
In order to devise the best signal acquisition strategy we implemented the simultaneous acquisition of the detector outputs by two different acquisition systems. According to Fig.~\ref{setup_TWB}, the electronic signal from the two SiPM detectors was split in two and sent to two synchronous boxcar-gated integrators (SR250, Stanford Research Systems) and to a desktop waveform digitizer (DT5720, CAEN) operating at 12-bit resolution and at 250-MS/s sampling rate. The boxcar integrator performs an analogical integration of the signal over gates of variable widths centered on the signal peak. For the comparison between the acquisition chains, we set the gate width of the boxcar at $\tau=50$~ns, while the output of the digitizer was integrated off-line over different time gates, $\tau=50, 100, 350$~ns. We note that the gate width can be pushed down to 10~ns for the boxcar, while to have a reasonable number of point in the integral, the minimum gate for the digitizer is 50~ns.\\
As a test of nonclassicality for the TWB state, we consider the noise reduction factor
\begin{equation}\label{NRF}
R = \frac{\sigma^2(n_1-n_2)}{\langle n_1 + n_2\rangle}\ ,
\end{equation}
where $\sigma^2(\cdot)$ is the variance of the difference in the number of photons on the two arms of the TWB measured shot-by shot. It is well known that the condition $R < 1$ is sufficient for nonclassicality.
For a multi-mode TWB state with real detection ($\mu_1 \neq \mu_2$, $\langle m_1 \rangle \neq \langle m_2 \rangle$), Eq.~(\ref{NRF}) takes the form \cite{arimondo}
\begin{equation}
R=1-2{\frac{\sqrt{\eta_s\langle m_s \rangle \eta_i \langle m_i \rangle}}{\langle m_s \rangle+\langle m_i \rangle}}+\frac{1}{\sqrt{\mu_s\mu_i}}\frac{(\langle m_s \rangle-\langle m_i \rangle)^2}{\langle m_s \rangle+\langle m_i \rangle},
\label{NRF2}
\end{equation}
which attains the limit $R = 1 - \eta$ under ideal conditions.
\begin{figure}[h!]
\begin{center}
\includegraphics[width=0.7\columnwidth]{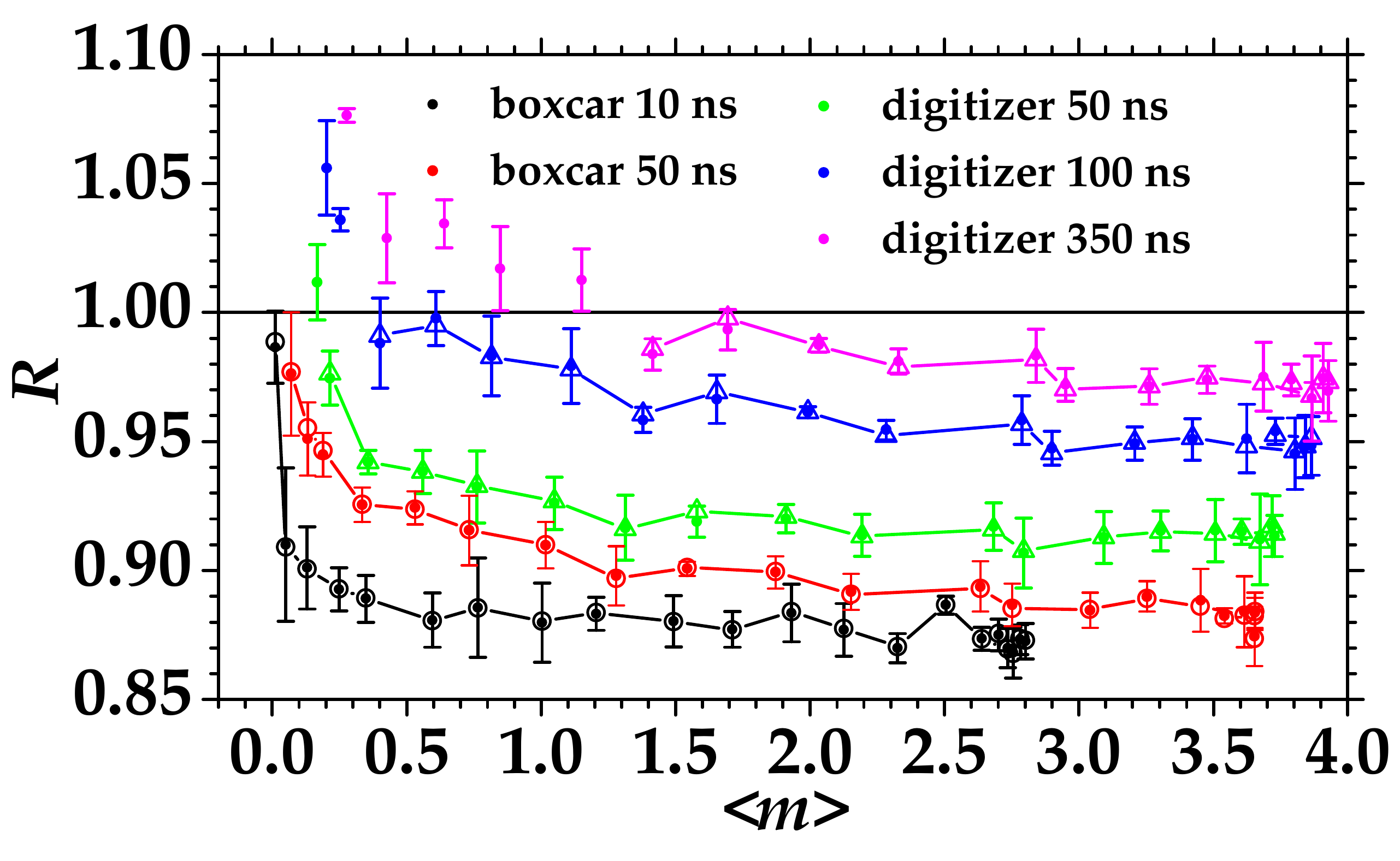}
\end{center}
\caption{Noise reduction factor as a function of the number of photons measured in a single arm. Values for different integration gates. Top to bottom: digitizer 350~ns (magenta), digitizer 100~ns (blue), digitizer 50~ns (green), boxcar 50~ns (red) and boxcar 10~ns (black). Full symbols are the experimental data and line+open symbols are the theoretical expectation evaluated according to Eq.~(\ref{NRF2}).}
\label{grnrf}
\end{figure}
\noindent
In Fig.~\ref{grnrf}, we plot the measured values of $R$ as a function of the mean number of photons measured in one of the two TWB arms. Different colors correspond to different gate widths (see caption for details). The data are well superimposed to the theoretical curves evaluated according to Eq.~(\ref{NRF2}), in which we use the measured values of $\langle m_j \rangle$, $\mu_j$ and $\eta_j$, with $j=1,2$ \cite{lamperti}. In particular, we used the measured values of $\langle m_1 \rangle$ and $\langle m_2 \rangle$. The number of modes $\mu$ was estimated by the first two moments of photon-number statistics, whereas $\eta_j$ was considered equal in the two arms, namely $\eta_s = \eta_i = \eta$, and obtained from the experimental values of the noise reduction factor setting $R = 1 - \eta$.\\
By comparing the data corresponding to the different gate widths set for the digitizer, we can argue that the shorter the gate the lower the values of $R$, a result consistent with the model of the detection response \cite{manuscriptARXIV}. Moreover, it is clear that the boxcar-gated integrator performs better than the digitizer for the same gate width (see the comparison for $\tau=50$~ns in Fig.~\ref{grnrf}($a$), red and green data). The reason for that is the finite sampling rate of the digitizer, which is not sufficient to reliably sample the very fast output trace in correspondence of the signal peak.
To shorten the acquisition gate to the limits, in a subsequent measurement we set the boxcar gate width to the minimum value compatible with the measurement stability, $i.e.$ $\tau=10$~ns. The resulting values of $R$, shown in Fig.~\ref{grnrf} as black symbols, are lower than the others, thus indicating an improved capability of detecting the nonclassicality of the TWB state. This last result indicates that the cleanest information on the detected light is contained in the peak of the electronic output signal. For this reason, we implemented a peak-and-hold circuit to detect the peak value of the electronic output signal \cite{manuscriptARXIV}. In Fig.~\ref{peak} we show preliminary results for the values of $R$ as a function of the mean number of photons measured with the peak-and-hold apparatus (red) compared to those measured with the boxcar at $\tau=10$~ns (black). We note that the use of the peak and hold improves the performance of the digitizer, but still gives worse results in comparison with the analogical boxcar-gated integrator. Further improvement of the peak-and-hold acquisition system is necessary to assess the limit of such technology.\\
Once demonstrated the nonclassical nature of the correlations between the parties of the TWB state, we can exploit it to prepare single-beam nonclassical states by conditioning operations \cite{laurat03,laurat04,ourj06,EPL,lamperti,Iskhakov16}. The conditioning protocol consists in choosing one value of the output, $m_{\mathrm{cond}}$, measured on one of the two parties of the TWB state, say the beam $\# 1$, and selecting the corresponding value $m_2$ on the other beam. In general, the conditioning operation modifies the statistics of the original state. In the present case, due to the perfect photon-number correlations in the TWB, for a perfect detection chain we would expect the conditional generation of the Fock state $|m_{\mathrm{cond}}\rangle$, but for real experimental schemes we can only obtain a sub-Poissonian state characterized by the following expression for the Fano factor
%
\begin{eqnarray}
F_{\mathrm{cond}} &=& (1-\eta)\left[1+\frac{1}{(\langle m_2\rangle+\mu)} \times\right. \nonumber\\
&\times &\left. \frac{\langle m_2\rangle (m_{\mathrm{cond}}+\mu)(\langle m_2\rangle +\eta \mu)} {(m_{\mathrm{cond}} +\mu)(\langle m_2\rangle +\eta \mu)-\eta\mu (\langle m_2\rangle+\mu)} \right]\ ,
\label{fanoCONDIZ}
\end{eqnarray}
%
where the overall detection efficiency $\eta$ can be experimentally obtained from the sub-shot-noise values displayed in Fig.~\ref{grnrf} and $\langle m_2\rangle$ is the mean value of the state $\# 2$ before conditioning. We note that $F_{\mathrm{cond}}<1$ for $m_{\mathrm{cond}}\geq 1$.
\begin{figure}[h!]
\begin{center}
\includegraphics[width=0.7\columnwidth]{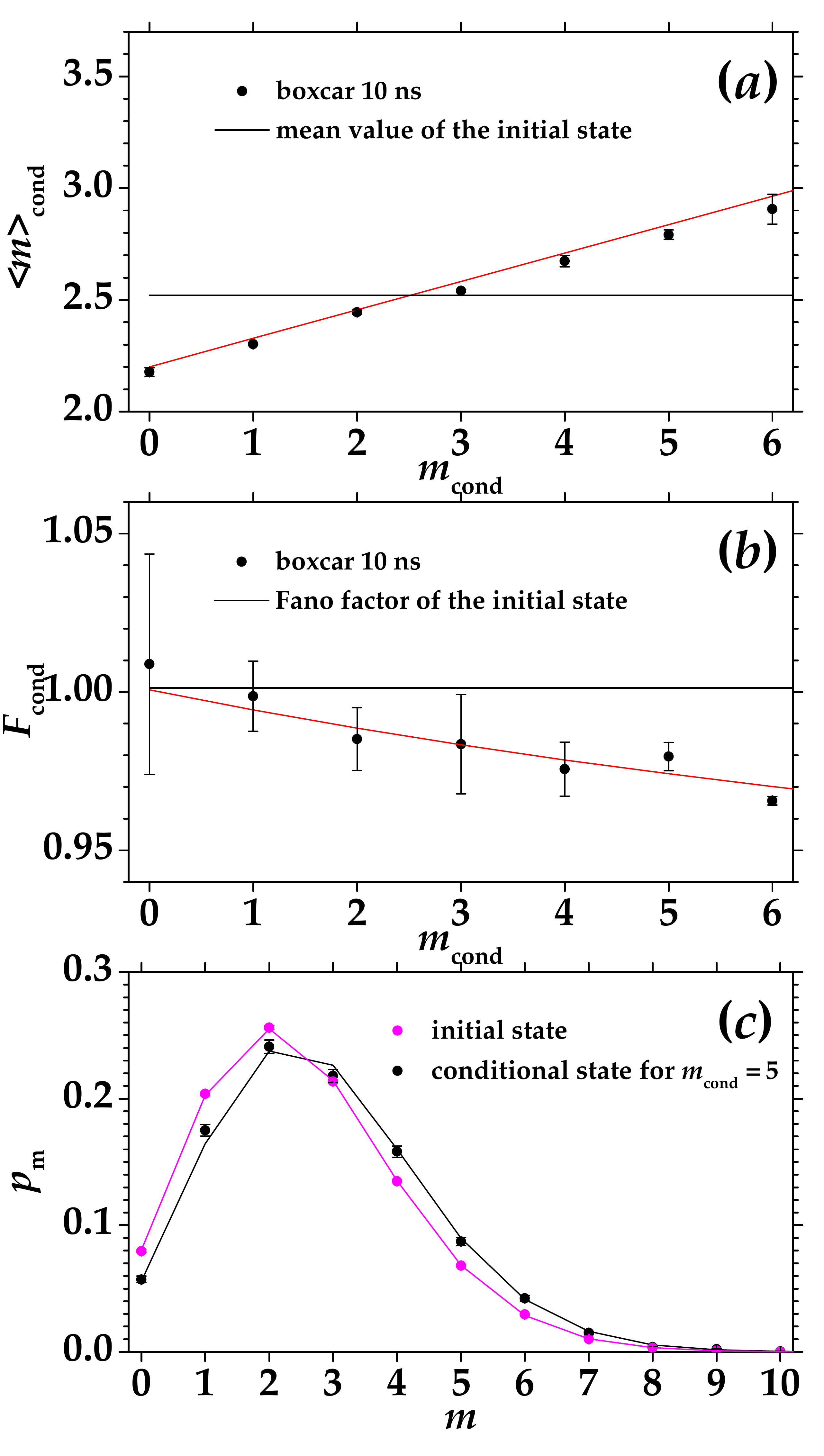}
\end{center}
\caption{($a$) Mean value of the conditional states at different conditioning values; ($b$) Fano factor of the conditional states; ($c$) Reconstructed photon-number statistics of unconditioned state (black dots) and conditional state (magenta dots) with $m_{cond} = 5$. The theoretical expectations are shown as black curves and well superimposed to the data.}
\label{conditional}
\end{figure}
In Fig.~\ref{conditional} ($a$), we plot the measured mean values of the conditional states generated by selecting different values of $m_{\mathrm{cond}}$ on beam $\# 1$ for signal acquisition with the boxcar integrator on 10~ns gate width, which is the best experimental condition we could achieve. The initial mean value of beam $\# 2$ is $\langle m_2\rangle=2.52$ (horizontal line in the figure). We note that, as expected, the mean value of the conditional state increases or decreases according to the conditioning value.  In Fig.~\ref{conditional} ($b$) we plot the measured Fano factor for the same conditional states shown in panel ($a$). We note that the values are below 1 and consistent with the theory evaluated according to Eq.~(\ref{fanoCONDIZ}) (red line). The horizontal line in the figure is at the value of the Fano factor of the original beam $\# 2$ ($F=1.00126$).
Finally, Fig.~\ref{conditional} ($c$)) shows the statistical distribution of the detected photons for the state obtained by conditioning at $m_{\mathrm{cond}} = 5$  (black dots) along with the original distribution on beam $\# 2$. Continuous lines in the figure are theoretical predictions \cite{EPL}.\\
Even if the value of sub-Poissonianity is quite small, nevertheless the effect is present and interesting: we have obtained a mesoscopic sub-Poissonian state ($\langle m\rangle_{\mathrm{cond}}=2.79$). We note that similar conditioning operations performed on the data acquired by the digitizer do not perform equally well, and that the best results are obtained for the shortest gate, as expected from the results in Fig.~\ref{grnrf}. This result seems to indicate that the cleaner information on the detected light is obtained by integrating the values around the peak of the output signal.\\
In conclusion, we have presented, for the first time to our knowledge, a detection chain based on Silicon photomultipliers that can be used to test nonclassicality of mesoscopic quantum states of light. First of all, we have shown that these detectors can reveal the presence of nonclassical correlations between the two parties of a multi-mode TWB state. Second, we have demonstrated that they can also be used to generate and reconstruct nonclassical photon-number distributions. We notice that the intensity regime of both the original TWB state and the conditional sub-Poissonian states are well beyond the so-called single-photon level and that the system can be operated at room temperature. Further investigations are ongoing to optimize the peak-and-hold apparatus in order to make the system as compact and portable as possible.  All such features make the technology of SiPMs useful in several contexts of Quantum Optics and Quantum Information.

\end{document}